\documentstyle[lcws2002]{article}
\input{psfig}

\def\be{\begin{equation}}
\def\ee{\end{equation}}
\def\bea{\begin{eqnarray}}
\def\eea{\end{eqnarray}}
\def\Re{{\cal R \mskip-4mu \lower.1ex \hbox{\it e}\,}}
\def\Im{{\cal I \mskip-5mu \lower.1ex \hbox{\it m}\,}}
\def\ie{{\it i.e.}}

\def\etal{{\it et al.}}

\def\sub#1{_{\lower.25ex\hbox{$\scriptstyle#1$}}}
\def\tev{\,{\ifmmode\mathrm {TeV}\else TeV\fi}}
\def\gev{\,{\ifmmode\mathrm {GeV}\else GeV\fi}}
\def\mev{\,{\ifmmode\mathrm {MeV}\else MeV\fi}}
\def\mpl{\ifmmode M_{pl}\else $M_{pl}$\fi}
\def\to{\rightarrow}

\def\subw{_{\rm w}}
\def\mh{\ifmmode m\sbl H \else $m\sbl H$\fi}
\def\mch{\ifmmode m_{H^\pm} \else $m_{H^\pm}$\fi}
\def\mt{\ifmmode m_t\else $m_t$\fi}
\def\mc{\ifmmode m_c\else $m_c$\fi}
\def\mz{\ifmmode M_Z\else $M_Z$\fi}
\def\mw{\ifmmode M_W\else $M_W$\fi}
\def\mws{\ifmmode M_W^2 \else $M_W^2$\fi}
\def\mhs{\ifmmode m_H^2 \else $m_H^2$\fi}   
\def\mzs{\ifmmode M_Z^2 \else $M_Z^2$\fi}
\def\mts{\ifmmode m_t^2 \else $m_t^2$\fi}
\def\mcs{\ifmmode m_c^2 \else $m_c^2$\fi}
\def\mchs{\ifmmode m_{H^\pm}^2 \else $m_{H^\pm}^2$\fi}
\def\ztwo{\ifmmode Z_2\else $Z_2$\fi}
\def\zone{\ifmmode Z_1\else $Z_1$\fi}
\def\mtwo{\ifmmode M_2\else $M_2$\fi}
\def\mone{\ifmmode M_1\else $M_1$\fi}
\def\tb{\ifmmode \tan\beta \else $\tan\beta$\fi}
\def\xw{\ifmmode x\subw\else $x\subw$\fi}
\def\ch{\ifmmode H^\pm \else $H^\pm$\fi}
\def\lum{\ifmmode {\cal L}\else ${\cal L}$\fi}
\def\inpb{\,{\ifmmode {\mathrm {pb}}^{-1}\else ${\mathrm {pb}}^{-1}$\fi}}
\def\infb{\,{\ifmmode {\mathrm {fb}}^{-1}\else ${\mathrm {fb}}^{-1}$\fi}}
\def\epem{\ifmmode e^+e^-\else $e^+e^-$\fi}
\def\ppb{\ifmmode \bar pp\else $\bar pp$\fi}
\def\bsg{\ifmmode B\to X_s\gamma\else $B\to X_s\gamma$\fi}
\def\bsll{\ifmmode B\to X_s\ell^+\ell^-\else $B\to X_s\ell^+\ell^-$\fi}
\def\bstt{\ifmmode B\to X_s\tau^+\tau^-\else $B\to X_s\tau^+\tau^-$\fi}
\def\lamt{\ifmmode \tilde\lambda\else $\tilde\lambda$\fi}
\def\shat{\ifmmode \hat s\else $\hat s$\fi}
\def\that{\ifmmode \hat t\else $\hat t$\fi}
\def\uhat{\ifmmode \hat u\else $\hat u$\fi}

\newskip\zatskip \zatskip=0pt plus0pt minus0pt

\begin{document}

\rightline{\vbox{\halign{&#\hfil\cr
SLAC-PUB-9492\cr
September 2002\cr}}}
\vspace{0.8in}

\title{EFFECTS OF RADION MIXING ON THE STANDARD MODEL HIGGS BOSON}

\author{ THOMAS G. RIZZO 
\footnote{Work supported by the Department of Energy, Contract 
DE-AC03-76SF00515}}

\address{Stanford Linear Accelerator Center,
Stanford,\\ CA 94309, USA}

\maketitle\abstracts{We discuss how mixing between the Standard Model 
Higgs boson and the radion of the Randall-Sundrum model can lead to 
significant shifts in the expected properties of the Higgs boson. In 
particular we show that the total and partial decay widths of the Higgs, 
as well as the $h\to gg$ branching fraction, can be substantially altered 
from their SM expectations, while the remaining branching fractions are 
modified less than about $5\%$ for most of the parameter space volume. 
Precision measurements of Higgs boson properties at at a Linear Collider 
are shown to probe a large region of the Randall-Sundrum model parameter 
space.}

The Randall-Sundrum (RS) model{\cite {big}} offers a potential solution to the 
hierarchy problem. In this model the Standard Model (SM) fields lie 
on one of two branes that are embedded in $AdS_5$  
described by the metric $ds^2=e^{-2k|y|}\eta_{\mu\nu}dx^\mu dx^\nu-dy^2$, 
where $k$ is the 5-d curvature 
parameter of order the Planck scale. To solve the hierarchy problem 
the separation between the two branes, $r_c$, must have a value of 
$kr_c \sim 11-12$. That this quantity can be stabilized and made natural has 
been demonstrated by a number 
of authors and leads directly to the existence of a radion($r$), 
which corresponds to a quantum excitation of the brane separation. It can be 
shown that the radion couples to the trace of the 
stress-energy tensor formed from the SM fields  with a strength $\Lambda$ of 
order the TeV scale, \ie, ${\cal L}_{eff}=-r~T^\mu_\mu /\Lambda$. 
This leads to couplings for the radion that 
are qualitatively similar to those of the SM 
Higgs boson. The radion mass ($m_r$), which is generated by the brane 
separation stabilization mechanism, is expected to be significantly 
below the scale $\Lambda$ implying that the radion may be 
the lightest new field predicted by the RS model. 

On general grounds of covariance, the radion may mix with the SM Higgs field 
through an interaction term of the form 
\begin{equation}
S_{rH}=-\xi \int d^4x \sqrt{-g_w} R^{(4)}[g_w] H^\dagger H\,,
\end{equation}
where $H$ is the Higgs doublet field, 
$R^{(4)}[g_w]$ is the Ricci scalar constructed out of the induced metric $g_w$
on the SM brane,  and $\xi$ is a dimensionless mixing parameter assumed to be 
of order unity and with unknown sign. The 
above action induces kinetic mixing between the `weak eigenstate' $r_0$ and 
$h_0$ fields which can be removed through a set of field redefinitions and 
rotations. Clearly, since the radion and Higgs boson 
couplings to other SM fields 
differ this mixing will induce modifications in the usual SM expectations for 
the Higgs decay widths and branching fractions.

To make unique predictions in this scenario we need to 
specify four parameters: the masses of the {\it physical} Higgs and radion 
fields, $m_{h,r}$, the mixing 
parameter $\xi$ and the ratio $v/\Lambda$, where $v$ is 
the vacuum expectation value of the SM Higgs $\simeq 246$ GeV. Clearly the 
ratio $v/\Lambda$ cannot be too large as $\Lambda$ is already bounded 
from below by collider and electroweak precision 
data; for definiteness we will take $v/\Lambda \leq 0.2$ and 
$-1 \leq \xi \leq 1$. 
When we require that the weak basis mass-squared parameters of the radion and 
Higgs fields be real, as is required by hermiticity, we obtain an additional 
constraints on the ratio of the physical radion and Higgs masses which only 
depends on the product $|\xi| {v\over {\Lambda}}$. These constraints are 
incorporated into the analysis below.

\begin{figure}[htbp]
\centerline{
\psfig{figure=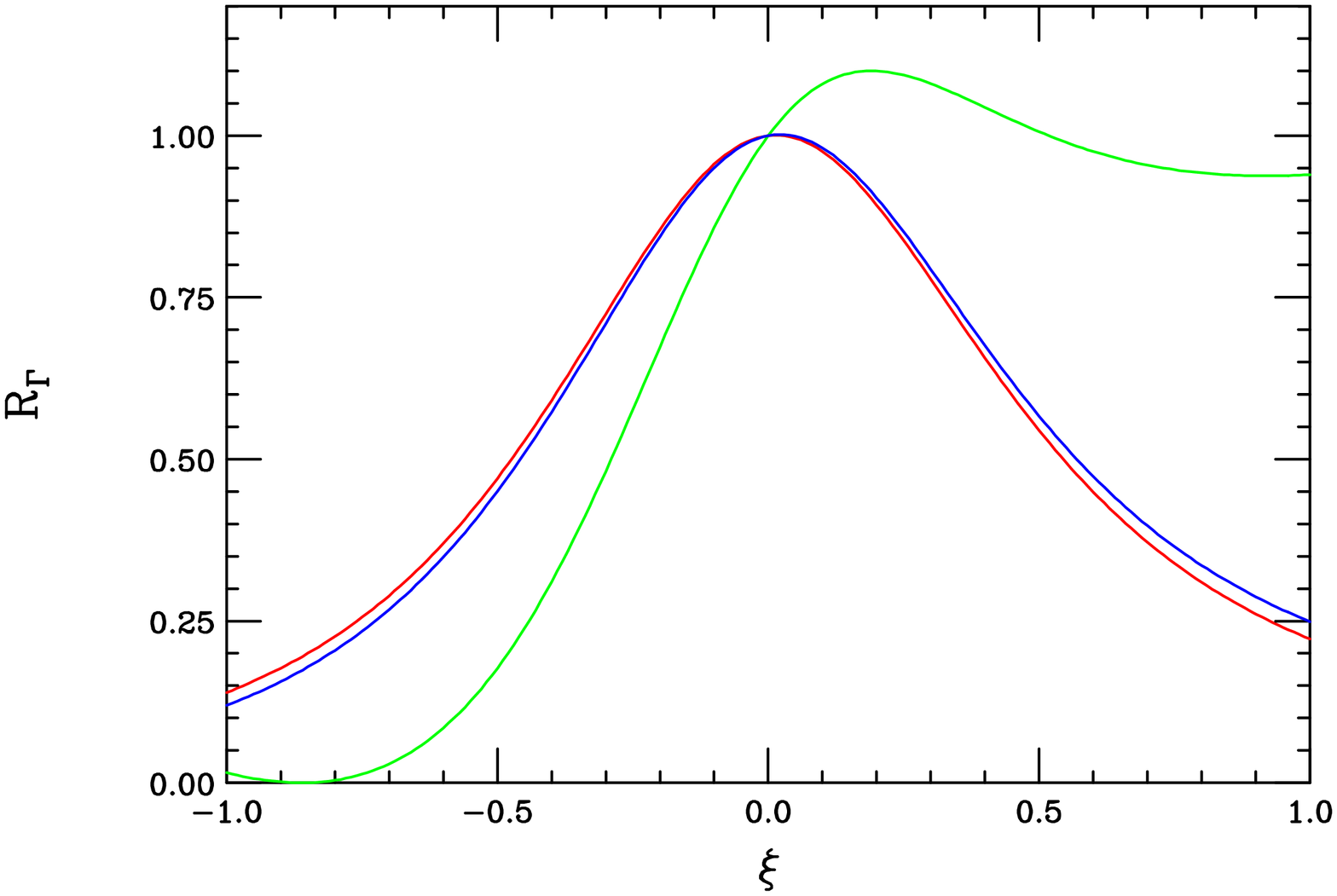,height=5cm,width=5.6cm,angle=90}
\hspace*{5mm}
\psfig{figure=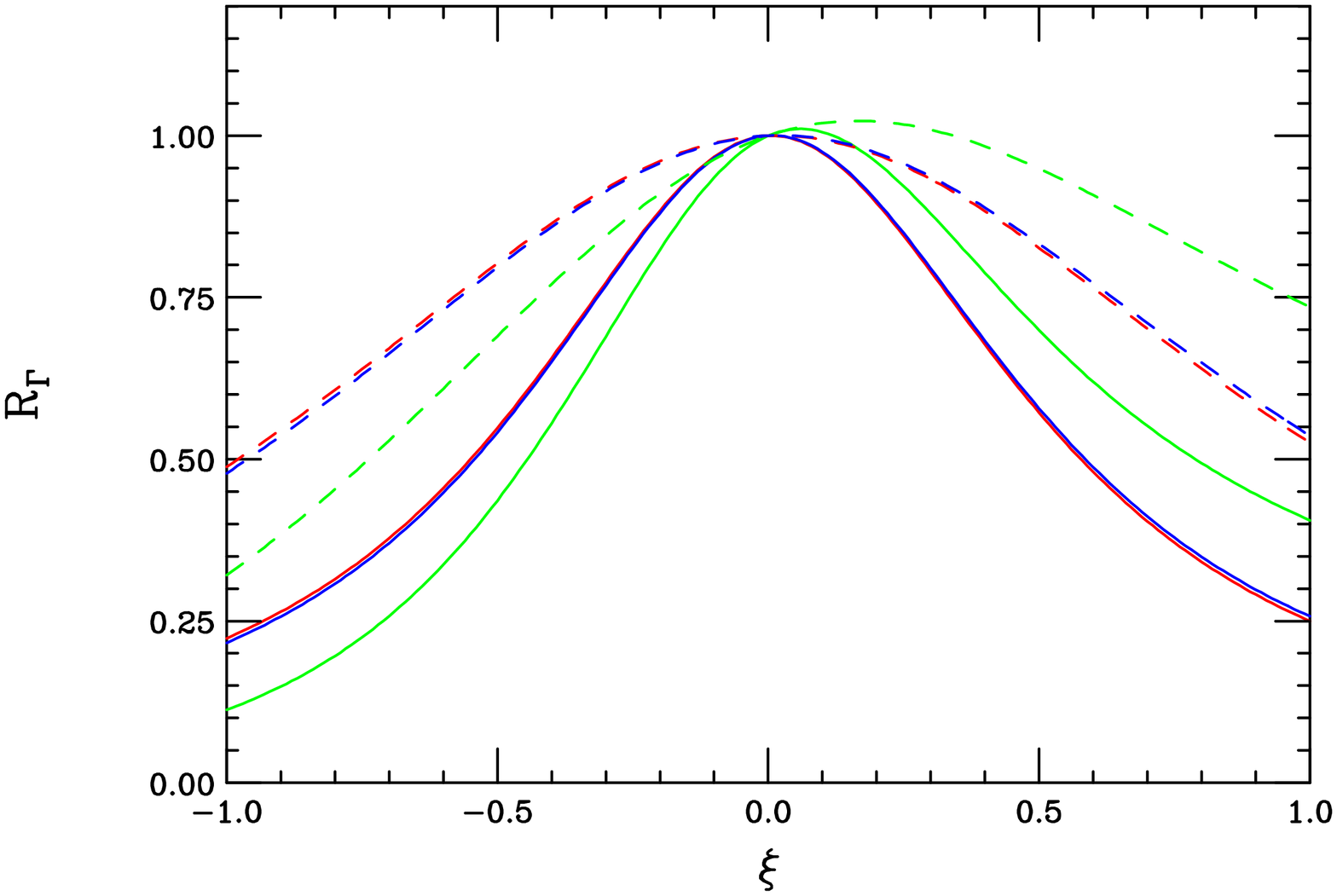,height=5cm,width=5.6cm,angle=90}}
\vspace*{0.1cm}
\caption{Ratio of Higgs widths to their SM values, $R_\Gamma$, as a function 
of $\xi$ assuming a physical Higgs mass of 125 GeV: red for fermion pairs or 
massive gauge boson pairs, green for gluons and blue for photons. In the left 
panel we assume $m_r=300$ GeV and $v/\Lambda=0.2$. In the right panel the 
solid(dashed) curves are for $m_r=500(300)$ GeV and $v/\Lambda=0.2(0.1)$.}
\label{fig2}
\end{figure}

Following the notation of Giudice \etal , the coupling of the 
physical Higgs to the SM fermions and massive gauge bosons $V=W,Z$ 
is given by
\begin{equation}
{\cal {L}}={-1\over {v}}(m_f\bar ff-m_V^2 V_\mu V^\mu)[\cos \rho \cos \theta +
{v\over {\Lambda}}(\sin \theta-\sin \rho \cos \theta)]h\,,
\end{equation}
where the angle $\rho$ is given above and $\theta$ can be calculated in 
terms of the 
parameters $\xi$ and $v/\Lambda$ and the physical Higgs and radion masses. 
Denoting the combinations $\alpha=\cos \rho \cos \theta$ and 
$\beta=\sin \theta-\sin \rho \cos \theta$, the corresponding Higgs 
coupling to gluons 
can be written as $c_g {\alpha_s\over {8\pi}}G_{\mu\nu}G^{\mu\nu}h$ with 
$c_g={-1\over {2v}}[(\alpha +{v\over {\Lambda}}\beta)F_g
-2b_3\beta {v\over {\Lambda}}]$ where $b_3=7$ is the $SU(3)$ $\beta$-function 
and $F_g$ is a well-known kinematic function of the ratio of masses of the  
top quark to the physical Higgs. 
Similarly the physical Higgs couplings to two photons is now given by 
$c_\gamma {\alpha_{em}\over {8\pi}}F_{\mu\nu}F^{\mu\nu}h$ where 
$c_\gamma={1\over {v}}[(b_2+b_Y)\beta {v\over {\Lambda}}-(\alpha 
+{v\over {\Lambda}}\beta)F_\gamma]$, where $b_2=19/6$ and $b_Y=-41/6$ are the 
$SU(2)\times U(1)$ $\beta$-functions and $F_\gamma$ is another well-known 
kinematic function of the ratios of the $W$ and top masses to the physical 
Higgs mass. (Note that in the simultaneous limits $\alpha \to 1,~\beta \to 0$ 
we recover the usual SM results.) From these expressions we can now compute  
the change of the various decay widths and branching fractions of the
SM Higgs due to mixing with the radion.  

Fig.~\ref{fig2} shows the ratio of the various Higgs widths in 
comparison to their SM expectations as functions of the parameter $\xi$ 
assuming that $m_h=125$ GeV with different values of $m_r$ and 
${v\over {\Lambda}}$. We see several features right away: ($i$) the shifts in 
the widths to $\bar ff/VV$ and $\gamma \gamma$ final states are very similar; 
this is due to the relatively large magnitude of $F_\gamma$ while the 
combination $b_2+b_Y$ is rather small. ($ii$) On the otherhand the shift for 
the $gg$ final state is quite different since $F_g$ is smaller than $F_\gamma$ 
and $b_3$ is quite large. ($iii$) For relatively light radions with a low 
value of $\Lambda$ the Higgs decay 
width into the $gg$ final state can come close to vanishing due to a strong 
destructive interference between the two contributions to the 
amplitude for values of $\xi$ near -1. ($iv$) Increasing the value of $m_r$ 
has less of an effect on the width shifts than does a decrease in the ratio 
${v\over {\Lambda}}$.

\begin{figure}[htbp]
\centerline{
\psfig{figure=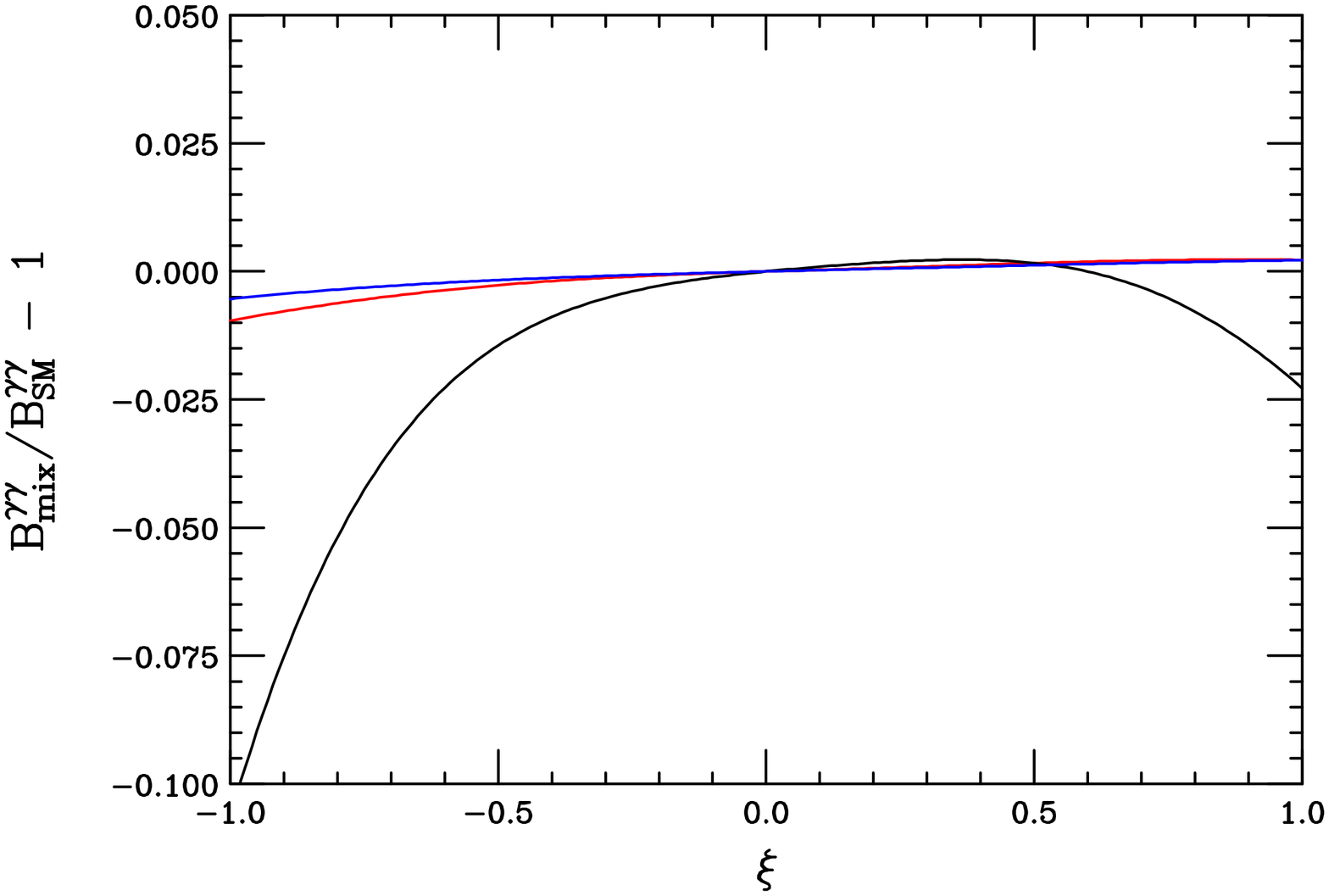,height=5cm,width=5.6cm,angle=90}
\hspace*{5mm}
\psfig{figure=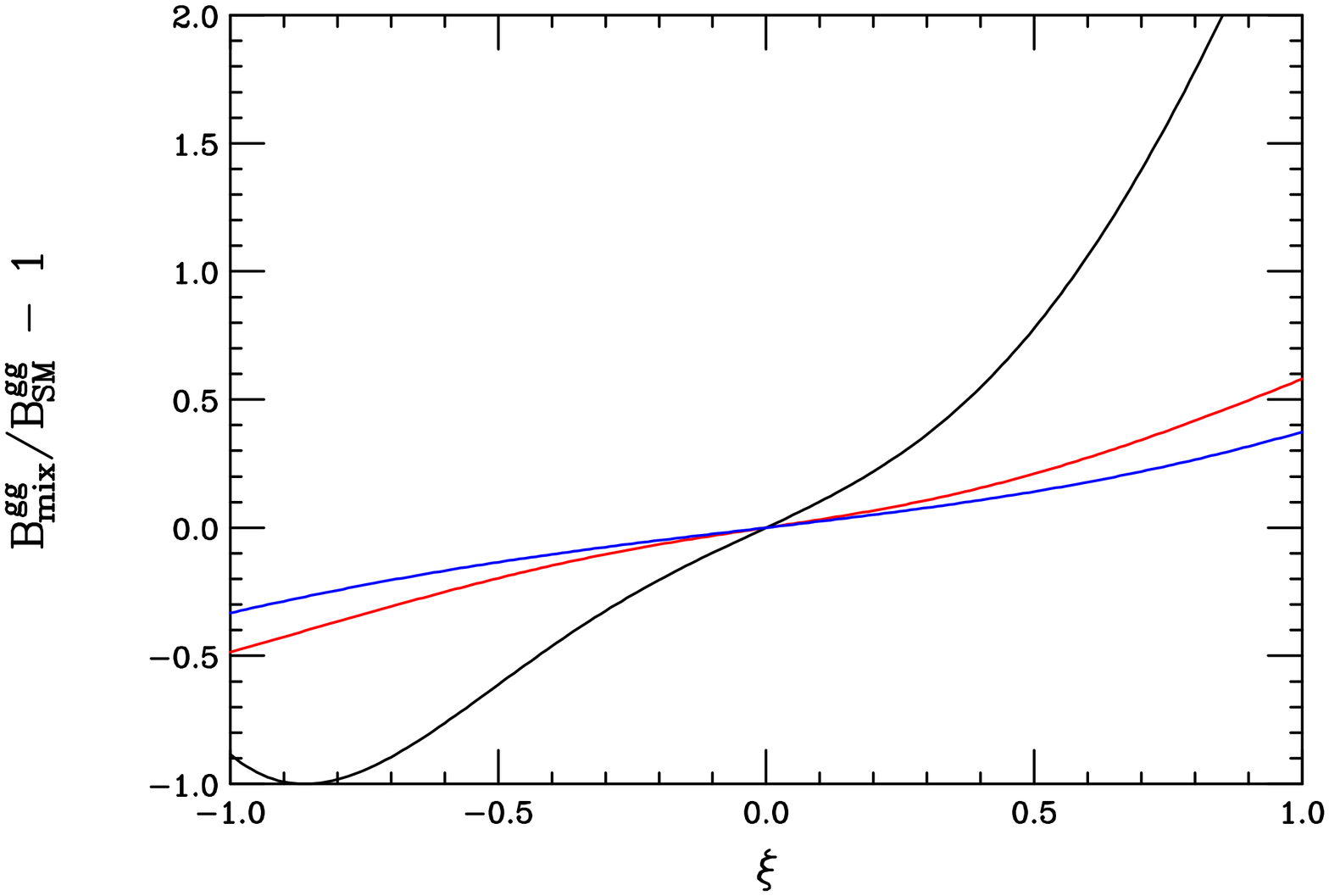,height=5cm,width=5.6cm,angle=90}}
\vspace*{0.1cm}
\centerline{
\psfig{figure=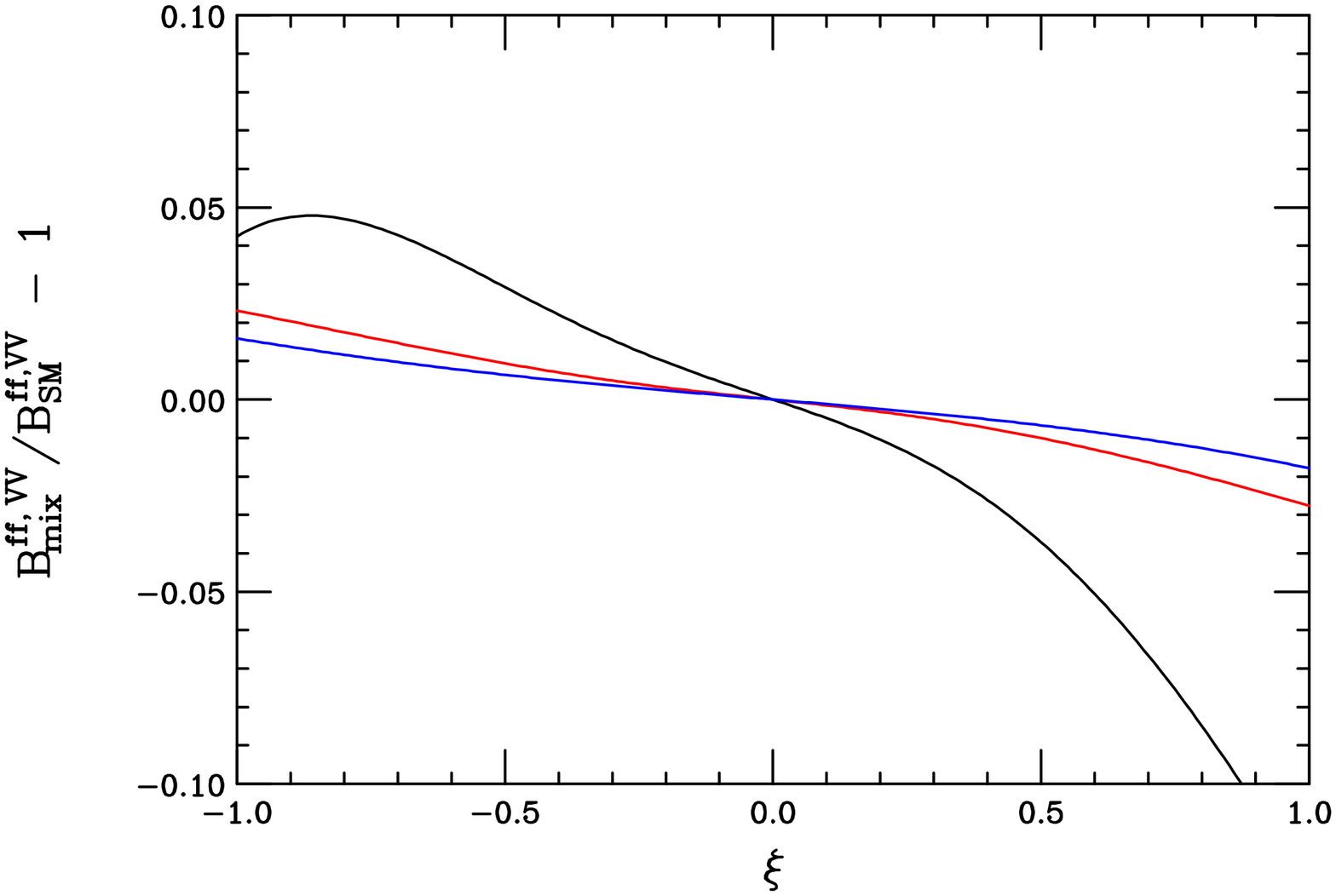,height=5cm,width=5.6cm,angle=90}
\hspace*{5mm}
\psfig{figure=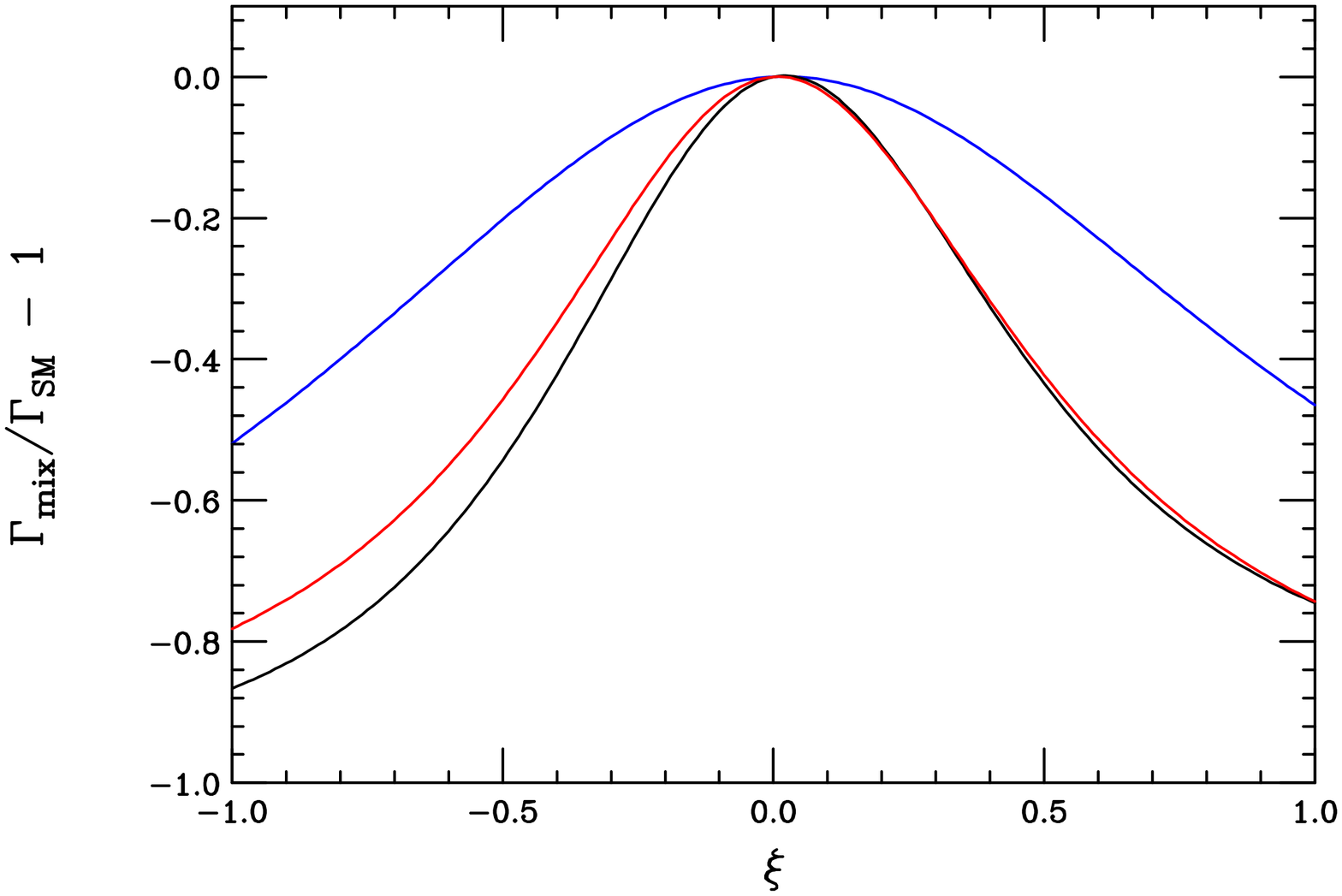,height=5cm,width=5.6cm,angle=90}}
\vspace*{0.1cm}
\caption{The deviation from the SM expectations for the Higgs
branching fraction into $\gamma\gamma$, $gg$, $f\bar f$, and $VV$ final
states as labeled, as well as for the total width.  The black, red, and
blue curves correspond to the parameter choices $m_r=300, 500, 300$ GeV
with $v/\xi=0.2, 0.2, 0.1$, respectively.}
\label{fig3}
\end{figure}

\begin{figure}[htbp]
\centerline{
\psfig{figure=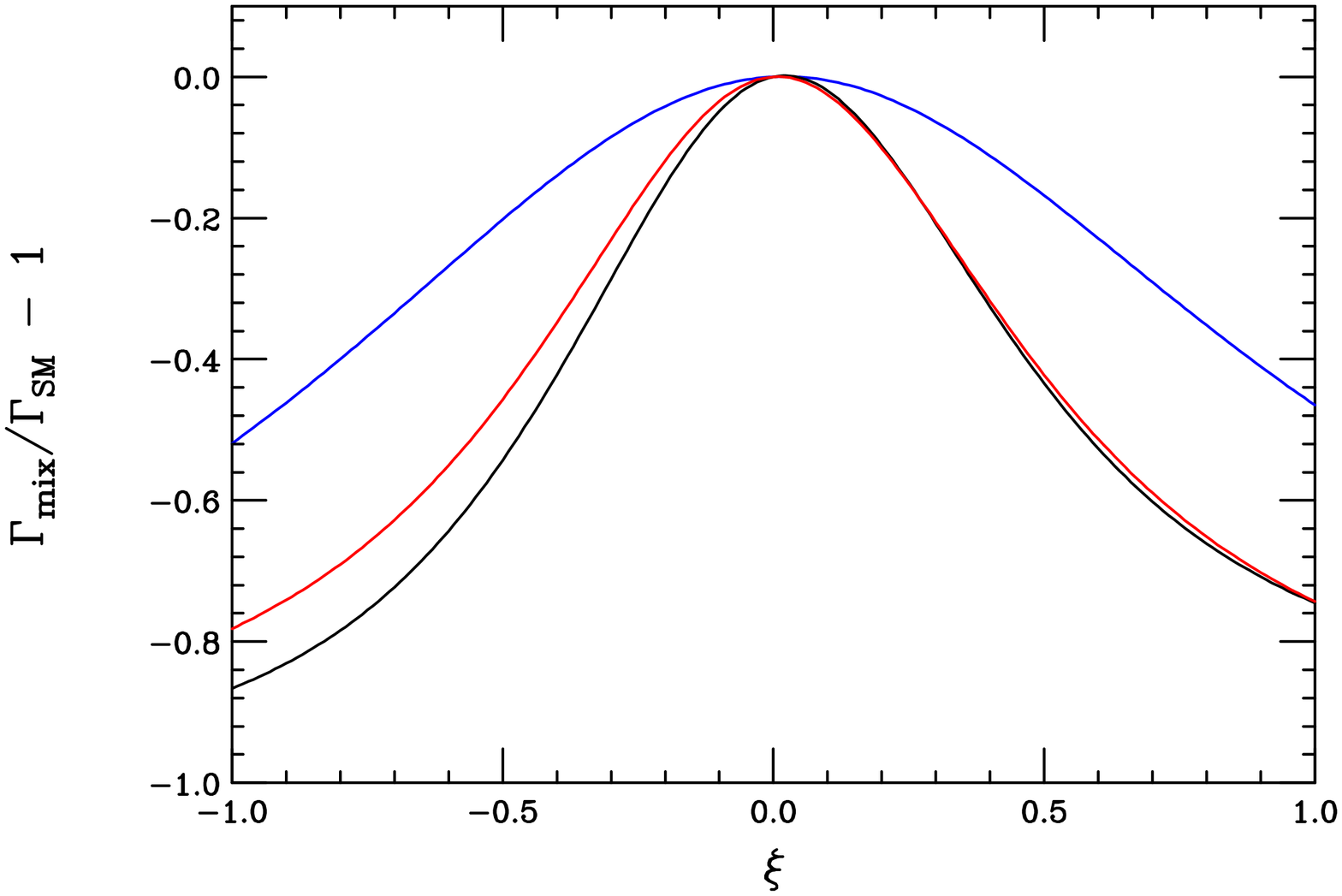,height=5cm,width=8.0cm,angle=90}}
\vspace*{0.1cm}
\caption{Lower bound on the mass of the Higgs boson from direct searches at 
LEP as a function of $\xi$ including the effects of mixing. The red 
(blue; green) 
curves correspond to the choice $m_r=300$ GeV, $v/\Lambda=0.2$
(500, 0.2; 300, 0.1).}
\label{fig4}
\end{figure}

The  deviation from the SM expectations for
the various branching fractions, as well as the total width, of the 
Higgs are displayed in Fig. \ref{fig3} as a function of the 
mixing parameter $\xi$.  We see that the gluon branching fraction and the
total width may be drastically different than that of the SM.  As we will 
see below the former
will affect the Higgs production cross section at the LHC.  However, the
$\gamma\gamma$, $f\bar f$, and $VV$, where $V=W,Z$ branching fractions
receive small corrections to their SM values, of order $5-10\%$
for almost all of the parameter region. Observation of these shifts 
will require the accurate determination
of the Higgs branching fractions obtainable at an $e^+e^-$ Linear 
Collider (LC) from which constraints on the radion model parameter 
space may be extracted as will be discussed below. These 
small changes in the $ZZh$ and $hb\bar b$ 
couplings of the Higgs boson can also lead to small reductions in the Higgs 
search reach from LEPII. This is shown in Fig.~\ref{fig4} for several sets of 
parameters; except for extreme cases this reduction in reach is rather 
modest. 

\begin{figure}[htbp]
\centerline{
\psfig{figure=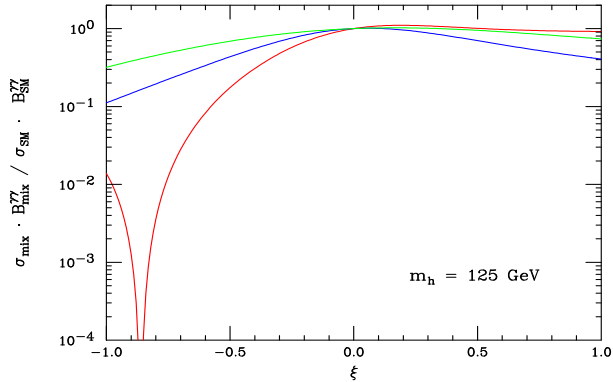,height=5cm,width=8.0cm,angle=90}}
\vspace*{0.1cm}
\caption{The ratio of production cross section times branching fraction
for $pp\to h\to\gamma\gamma$ via gluon fusion with radion mixing to the
SM expectations as a function of $\xi$.  The Higgs mass is taken to be 
125 GeV.  The red (blue; green) 
curves correspond to the choice $m_r=300$ GeV, $v/\Lambda=0.2$
(500, 0.2; 300, 0.1).}
\label{fig5}
\end{figure}

At the LHC the dominant production mechanism/signal for the light Higgs boson 
is via the gluon-gluon fusion through a triangle graph with subsequent decay 
into $\gamma \gamma$. Both the production cross section and the subsequent 
$\gamma \gamma$ branching fraction are modified by mixing as shown in 
Fig.~\ref{fig5}. This figure shows that the Higgs production rate in this mode 
at the LHC is always reduced in comparison to the expectations of the SM due 
to the effects of mixing. For some values of the parameters this reduction can 
be by more than an order of magnitude which could seriously hinder Higgs 
discovery via this channel at the LHC. 

\begin{figure}[htbp]
\centerline{
\psfig{figure=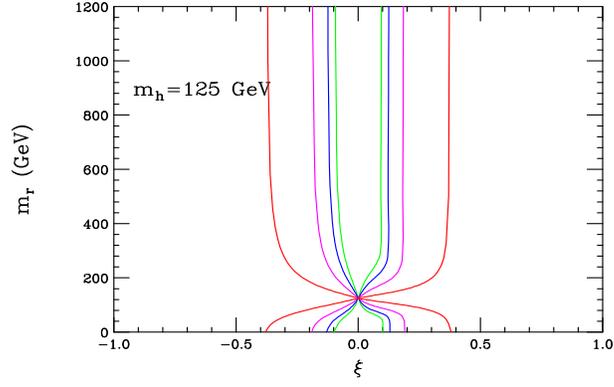,height=5cm,width=8.0cm,angle=90}}
\vspace*{0.1cm}
\caption{$95\%$ CL indirect bounds on the mass and mixing of the radion for 
a Higgs boson of mass 125 GeV arising from precision measurements of the 
Higgs couplings at the LHC and LC. The allowed region lies between the 
corresponding vertical pair of curves. From inner to outer the curves 
correspond to values of $v/\Lambda=0.2$, 0.15, 0.10, and 0.05, 
respectively.}
\label{fig6}
\end{figure}

Once both data from the LHC and LC become available the radion parameter 
space can be explored using both direct as well as indirect searches. For 
example, as discussed above, precision measurements of the Higgs boson 
couplings at these machines can be used to constrain the model parameter space 
beyond what may be possible through direct searches only. For purposes of 
demonstration let us assume that the LHC/LC measure these couplings to be 
consistent with the expectations of the SM. We then can ask what regions in 
the $\xi-m_r$ plane would remain allowed in this case as the ratio $v/\Lambda$ 
is varied; the results are shown in Fig.~\ref{fig6} assuming a Higgs mass of 
125 GeV. Here we see that if 
such a set of measurements were realized a large fraction of the parameter 
space would be excluded. Direct searches at LHC/LC would completely cover the 
lower portion of the remaining parameter space shown in the figure 
leaving only the high mass 
radion as a possibility under these circumstances.

\begin{figure}[htbp]
\centerline{
\psfig{figure=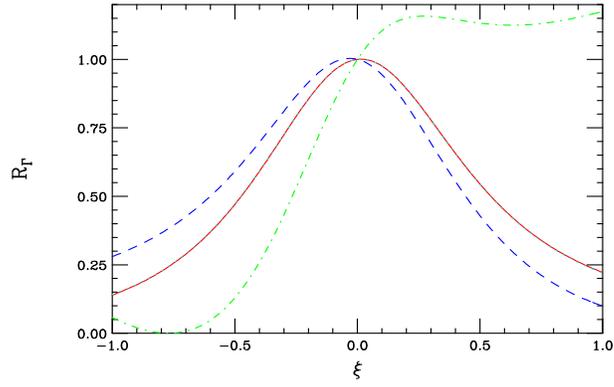,height=5cm,width=8.0cm,angle=90}}
\vspace*{0.1cm}
\caption{The effect of mixing on the partial widths of a 125 GeV Higgs 
boson, described by the parameter $\xi$, assuming $v/\Lambda=0.2$ and a 
radion mass of 300 GeV as 
discussed in the text. The solid(dash-dotted, dashed, dotted) corresponds to 
the $ZZ/W^+W^-(gg, \gamma\gamma, \bar ff)$ final states.}
\label{fig7}
\end{figure}

So far we have only considered the case where the SM fermion are confined to 
the TeV brane. It is possible instead to place the SM gauge and fermion fields 
in the RS bulk  which can lead to alterations in the radion  
couplings to these fields. Mixing with the Higgs could then lead to variations 
somewhat different than those discussed above. This possibility has been 
investigated and yields the results in Fig.~\ref{fig7} originates. 
Here we see the results analogous to those shown in the left panel of 
Fig. ~\ref{fig2}; note that in the case of bulk fields the expectations of 
the partial width shifts for fermions and massive vector fields are no longer 
degenerate. Qualitatively, however, the overall shifts in the Higgs boson 
couplings due to its mixing with the radion are found to be insensitive to 
whether or not the SM fields are in the RS bulk.

In summary, we see that Higgs-radion mixing, which is present in some
extra dimensional scenarios, can have a substantial effect on the
properties of the Higgs boson.  These modifications affect the widths
and branching fractions of Higgs decay into various final states, which
in turn can alter the expectations for Higgs production at both LEP and the 
LHC. For some regions of the parameters the size of these width and branching 
fraction shifts may require the precision of a Linear Collider to study in 
detail.

\end{document}